\documentclass{elsart}
\usepackage{float}
\usepackage{graphicx}
\usepackage{amssymb,amsfonts}

\newcommand{\ppLL}{\ensuremath{\bar\mathrm{p}\mathrm{p}\to\overline{\Lambda}\Lambda}}
\def\RE{\mathop{\Re{\rm e}}\nolimits}
\def\IM{\mathop{\Im{\rm m}}\nolimits}
\newcommand{\SLJ}[3]{\ensuremath{ {}^{#1}\mathrm{#2}_{#3}}}
\catcode`\@=11
\def\eqalign#1{\null\,\vcenter{\openup\jot\m@th
\ialign{\strut\hfil$\displaystyle{##}$&$\displaystyle{{}##}$\hfil
     \crcr#1\crcr}}\,}
\catcode`\@=12
\begin{document}
\begin{frontmatter}
\title{
\hspace*{10cm}{\normalfont\small ISN-03-11}\hfill\\[-8pt]%
\hspace*{10cm}{\normalfont\small nucl-th/0304015}\hfill\\[10pt] 
Constraints on spin observables in  \boldmath$\ppLL$\unboldmath\thanksref{LEAP03}}
\thanks[LEAP03]{Presented by J.M.R. at the LEAP03 Conference, Yokohama, Japan, March 2003, to appear in the Proceedings to be published as a special issue of \textsl{Nuclear Instrumentation and Methods in Physics Research}, ed.\ R.~Hayano et al.}
%\title{Constraints on spin observables in  \boldmath$\ppLL$\unboldmath}
\author[label1]{Jean-Marc Richard}, \author[label2]{Xavier Artru} 
 \address[label1]{ Institut des Sciences Nucl\'eaires, Universit\'e Joseph Fourier--CNRS-IN2P3\\
 53, avenue des Martyrs, 38026 Grenoble cedex, France}
  \address[label2]{Institut de Physique Nucl\'eaire de Lyon, Universit\'e Claude Bernard--CNRS-IN2P3\\
 4, rue Enrico Fermi, 69622 Villeurbanne cedex, France}
\begin{abstract}
We discuss model-independent inequalities for the spin observables in the reaction \ppLL, which have been measured at CERN with a polarised proton target.
\end{abstract}
\end{frontmatter}
% main text
%\section{Introduction}
%
Spin observables, when carefully analysed, give access to reaction mechanisms.
For instance, accurate measurements of proton--proton scattering  have made it possible to separate the  \SLJ3P0, \SLJ3P1, and \SLJ3P2\ phase-shifts.
At very low energy,  their ordering is  typical of tensor forces induced by  pion exchange. As energy increases, their behaviours are rather different,  revealing  a coherent spin-orbit force mediated by scalar and vector exchanges. 
The fine structure of protonium and the spin dependence of $\mathrm{\bar{p}p}$ scattering has been investigated at LEAR (for a recent review, see Ref.~\cite{Klempt:ap}). Antiproton--nucleus experiments also give valuable information. For instance the inelastic reaction $\mathrm{\bar{p} A\to \bar{p} A^*}$,
where $\mathrm{A}^*$ is a well identified excited level of the nucleus $\mathrm{A}$ with known quantum numbers, filters specific components of the antiproton interaction.

The  strangeness-exchange reaction $\ppLL$ \cite{Klempt:ap,Kingsberry,Alberg:fv} has been studied by the PS185 collaboration. The first data are well reproduced both by models with  K, K$^*$ exchange and by quark models \cite{Alberg:fv}. 
In particular, the dominance of the spin-triplet contribution over the spin-singlet contribution is reproduced in  both approaches. A final run with a polarised target was carried out in order to better identify the underlying mechanisms.

Many measurements are needed for a full reconstruction of  the amplitudes (up to an overall phase). Without enough information, ambiguities remain. On the other hand, when many observables are measured independently, one should address the problem of their compatibility. This is the subject of this note. 

In a situation like $\pi$N scattering, with two amplitudes $a$ and $b$, and spin observables
\cite{Leader:gr}
\begin{equation}
A=(|a|^2-|b|^2)/I_{0}~,\quad B=2\RE[ab^*]/I_{0}~,
\quad C=2\IM[ab^*]/I_{0}~,
\end{equation}
together with  $I_{0}=|a|^2+|b|^2$, both $A$ and $B$, as well as the sign of $C$ are needed to reconstruct the S-matrix. A value $A=0$, for instance, indicates that $|b|=|a|$, but does not inform about the relative phase of the two amplitudes.
If, instead, it happens that $A=+1$, then $b=0$, and the measurement of $B$ or $C$ is not necessary, except for cross-checking. In any case, the observables fulfil $A^2+B^2+C^2=1$, and thus $A^2+B^2\le1$, etc. 

Needless to say, the algebra becomes more complicated when two fermions are involved.
There are six independent amplitudes for \ppLL, once symmetries are enforced. The spin observables (times the differential cross section $I_{0}$) are expressed by quadratic relations of the type
\begin{equation} 
\label{Observables}
\eqalign{
 I_0&=|a|^2+|b|^2+|c|^2+|d|^2+|e|^2+|g|^2~,\cr
 P_{n}I_0&=2\RE (a{e}^*)+2\IM (d{g}^*)~,\cr
 A_{n}I_0&=2\RE (a{e}^*)-2\IM (d{g}^*),\cr
C_{nn}I_0&=|a|^2-|b|^2-|c|^2+|d|^2+|e|^2+|g|^2~,\cr
D_{nn}I_0&=|a|^2+|b|^2-|c|^2-|d|^2+|e|^2-|g|^2~,\cr
K_{nn}I_0&=|a|^2-|b|^2+|c|^2-|d|^2+|e|^2-|g|^2~,\quad\hbox{etc.}%
}
\end{equation}
Identities among these observables have been derived. For a review and references to the original papers, see, e.g., \cite{Leader:gr}.
With those identities in hand, one can deduce a number of constraining inequalities.
Richard and Elchikh \cite{RE}
 have studied the inequalities relating pairs of \ppLL\ observables in a empirical but systematic way. A powerful method is based on ``the generalised density matrix''. The positivity of this density matrix gives a number of inequalities involving sub-determinants. More details will be given elsewhere \cite{Artru:2003}. It should be noted that the framework developed for hadronic reactions can be applied to the transverse-momentum dependent helicity and transversity quark distributions in the nucleons. Existing inequalities  \cite{Structure} can be recovered.

\begin{figure}[H]
\begin{center}\includegraphics[width=.4\textwidth]{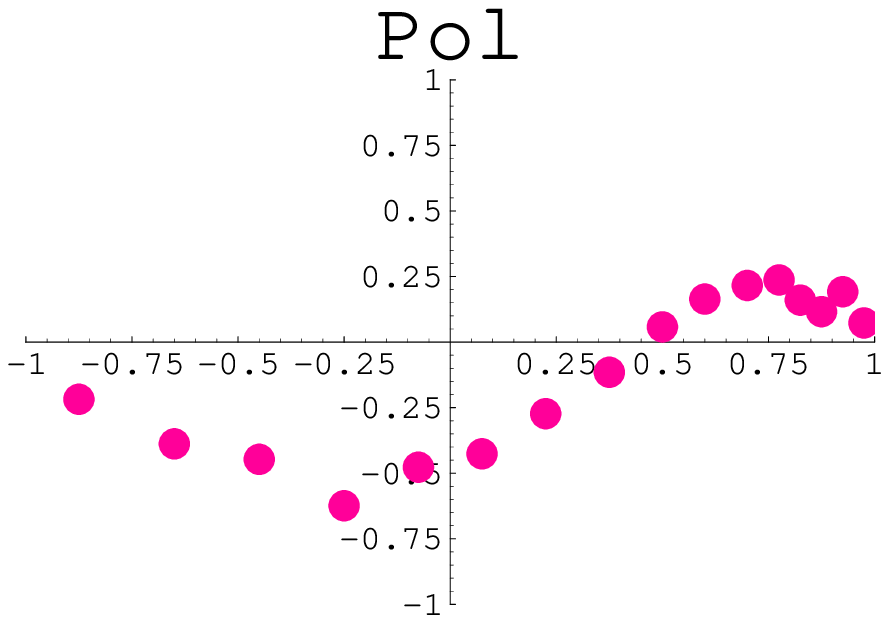}\
\ \includegraphics[width=.4\textwidth]{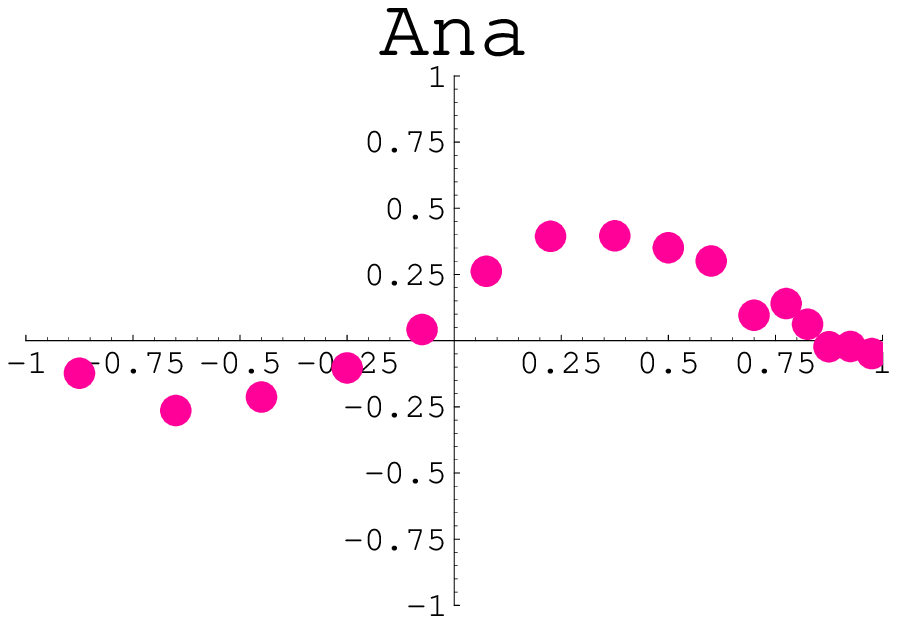}\end{center}
\caption{\label{Pol-Ana}Results of PS185 on polarisation and analysing power, as of function of $\cos\vartheta_{\mathrm{cm}}$}\end{figure}

For individual observables, the normalisation is such
$-1\le\mathcal{O}_i\le+1$. This can be checked on the
 published or preliminary data \cite{Kingsberry,Paschke,Bassalleck:2002sd}, as seen in Fig.~\ref{Pol-Ana}.

\begin{figure}[H]
\centering{%
\includegraphics[width=.4\textwidth]{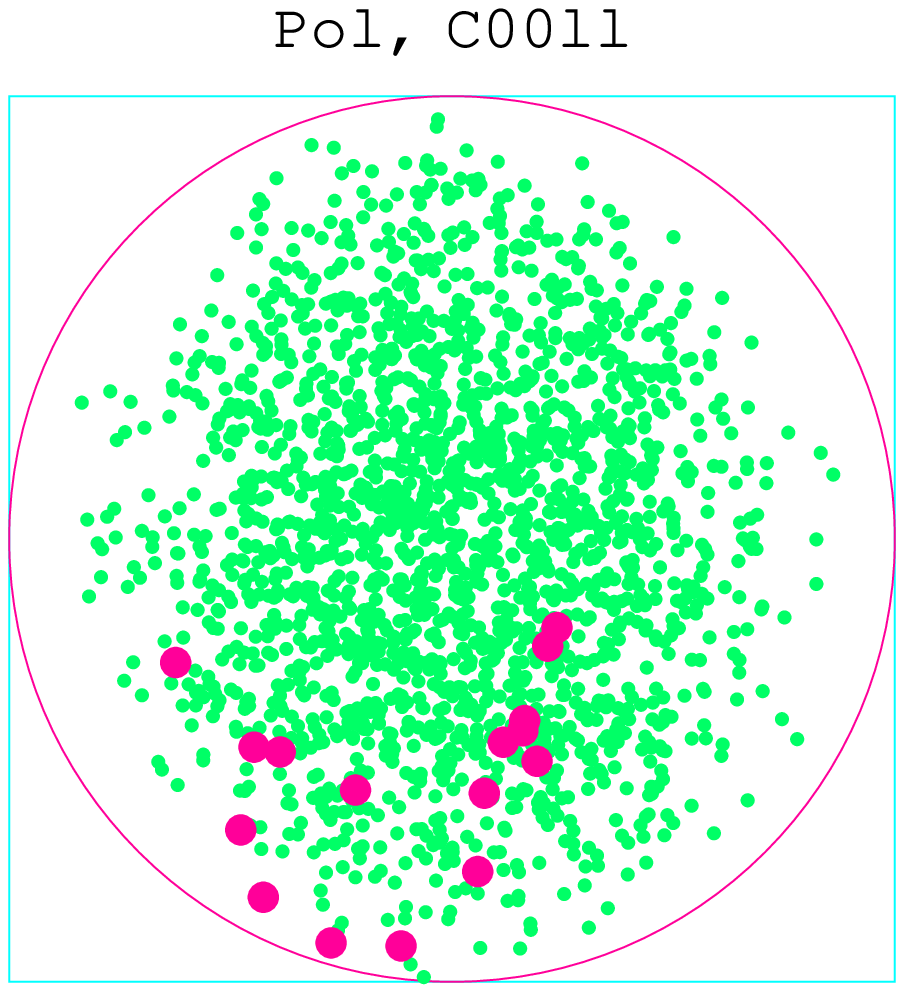}\hspace{.1\textwidth}%
\includegraphics[width=.4\textwidth]{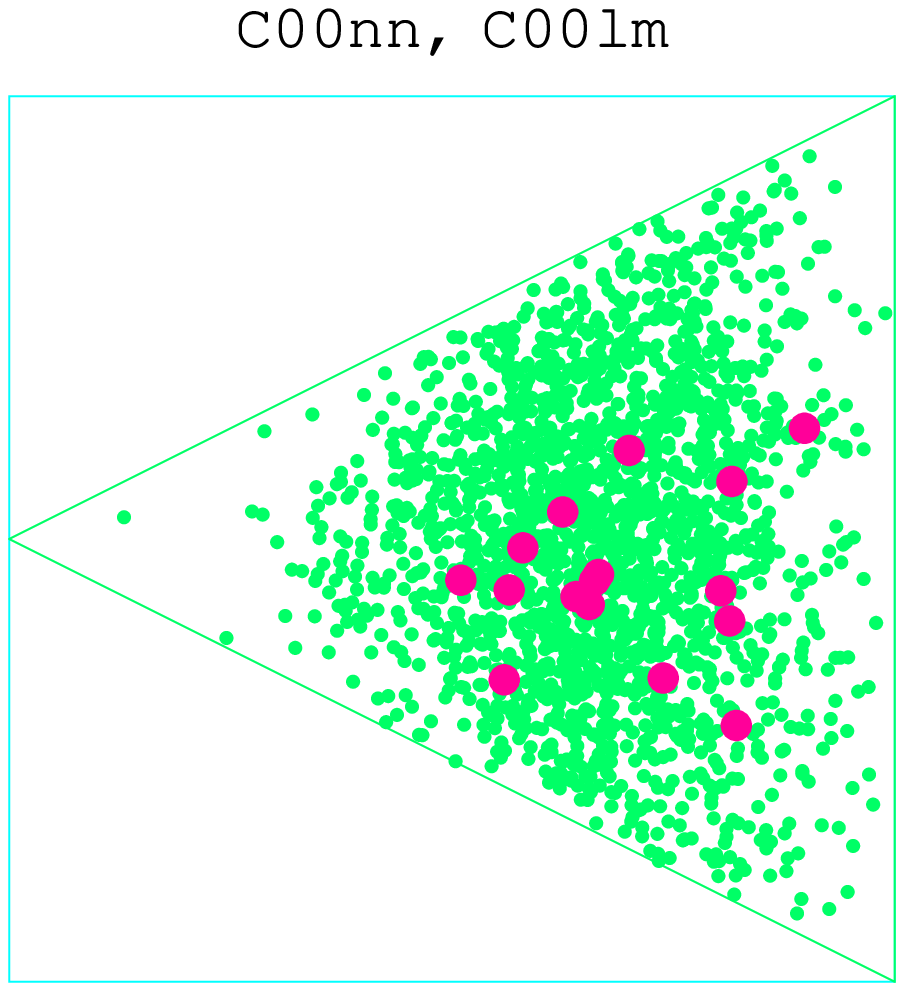}}
\caption{\label{Two} Pair of observables   restricted to the unit disk (left, here  polarisation and $C_{ll}$ are shown) or to a triangle (right, here $C_{nn}$ and $C_{lm}$ are shown). The small dots  correspond to unphysical, randomly generated, amplitudes, the larger dots, to actual data.}
\end{figure}

For pairs of observables, the domain is naively the square $\{ -1\le\mathcal{O}_i\le1,
-1\le\mathcal{O}_j\le1\}$. It is found that many (not all) pairs are restricted to a smaller area, an unit disk
$\mathcal{O}_i^2+\mathcal{O}_j^2\le1$
or  the inner part of a triangle
$\mathcal{O}_i^2-(1+\mathcal{O}_j)^2\le0$, as shown in Fig.~\ref{Two}.

\begin{figure}[H]
\centering{%
\includegraphics[width=.4\textwidth]{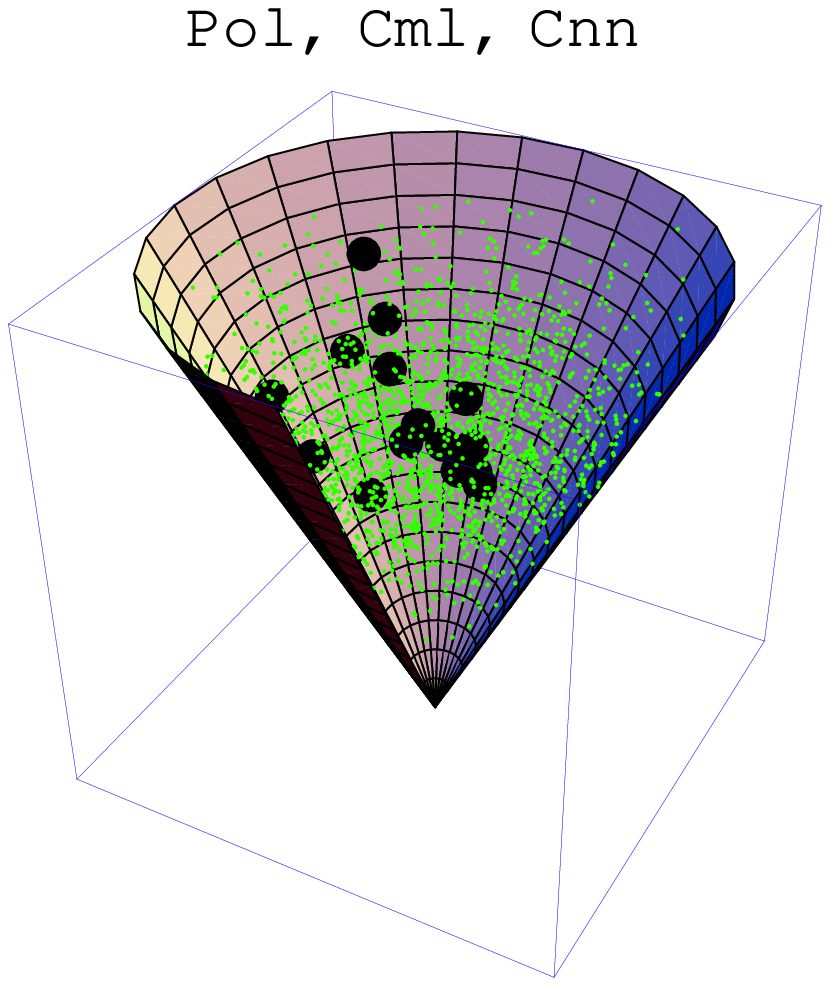}\hspace{.1\textwidth}%
\includegraphics[width=.4\textwidth]{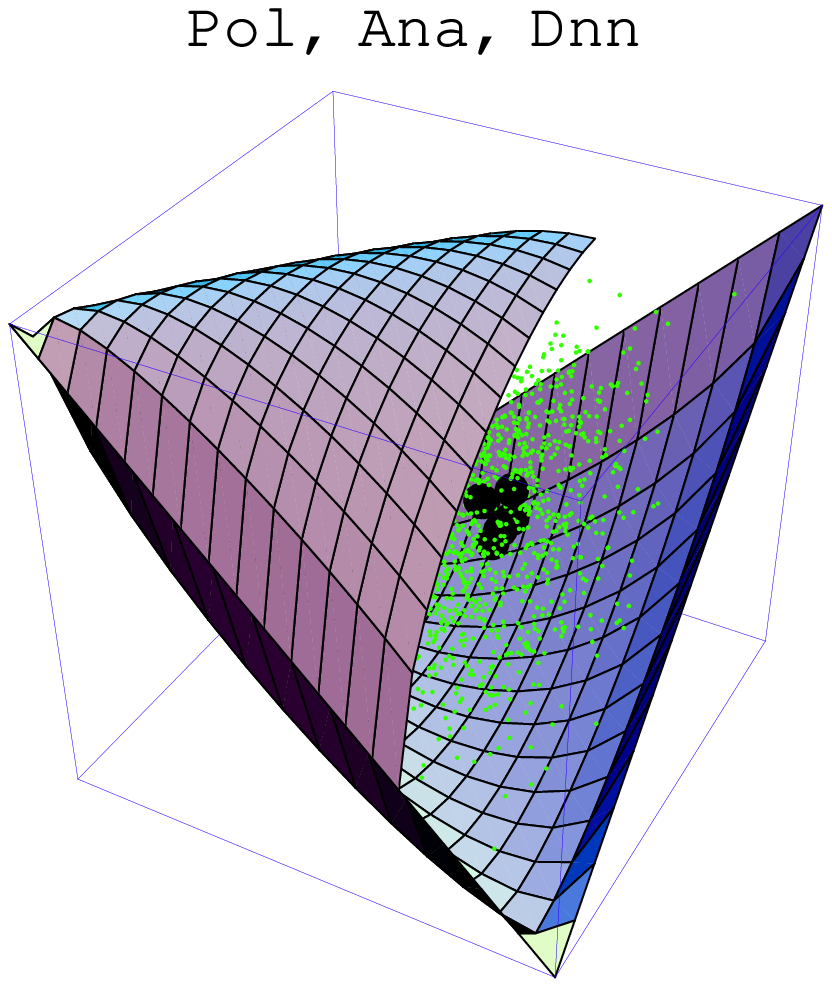}}
\caption{\label{Three} Triplet of observables restricted to the inner volume a cone or of a cubic. The small dots correspond to randomly generated amplitudes, the larger ones (partly hidden)  to actual data.}
\end{figure}

For  triplets of observables, three are many possibilities, among which the inner part of the unit sphere, the inner part of a cone (Fig.~\ref{Three}, left) and  the volume inside a symmetric cubic (right), which looks like a twisted cushion. This latter case is more interesting, as there is no restriction on any pair of the involved observables.
\textbf{Acknowledgments:}
J.M.R. would like to thank the organisers of LEAP2003 for the enjoyable and stimulating atmosphere of the Conference.
%

% \appendix

% %\section{}
% \label{}

%
\end{document}